\documentclass[tradiabstract]{aa}
\usepackage[dvipdf]{graphicx}
\usepackage{natbib}
\usepackage{txfonts}
\input{psfig}

\newcommand{\HI}{{\ion{H}{1}}}

\newcommand{\kms}{$\,$km$\,$s$^{-1}$}

\newcommand{\mJybeam}{mJy beam$^{-1}$}
\newcommand{\msun}{{$M_\odot$}}

\def\HI{H{\,\small I}}

\def\OIII{[O{\,\small III}]}
\def\OIV{[O{\,\small IV}]}
\def\emph#1{{\sl #1}}
\newcommand{\ltsima} {$\; \buildrel < \over \sim \;$}
\newcommand{\gtsima} {$\; \buildrel > \over \sim \;$}
\newcommand{\lta} {\lower.5ex\hbox{\ltsima}}
\newcommand{\gta} {\lower.5ex\hbox{\gtsima}}
\newcommand{\um }{$\mu$m}

\input{psfig}
\begin{document}
\title{PKS~1814-637: a powerful radio AGN  in a disk galaxy}
\titlerunning{The galactic environment of PKS~1814-637}
\authorrunning{Morganti et al.}
\author{R. Morganti\inst{1,2}, J. Holt\inst{3}, C. Tadhunter\inst{4},  C. Ramos Almeida\inst{4},  D. Dicken\inst{5}, K. Inskip\inst{6},  T. Oosterloo\inst{1.2}, T. Tzioumis\inst{7} }

\institute{Netherlands Foundation for Research in Astronomy, Postbus 2,
NL-7990 AA, Dwingeloo, The Netherlands
\and
Kapteyn Astronomical Institute, University of Groningen, P.O. Box 800,
9700 AV Groningen, The Netherlands
\and
Leiden Observatory, Leiden University, PO Box 9513, 2300 RA Leiden, The Netherlands. 
\and
Dep. Physics and Astronomy,
University of Sheffield, Sheffield, S7 3RH, UK
\and
Department of Physics and Astronomy, Rochester Institute of Technology, 84 Lomb Memorial Drive, Rochester 
NY 14623, USA 
\and
Max-Planck-Institut f\"ur Astronomie, K\"onigstuhl 17, D-69117 Heidelberg, Germany
\and 
CSIRO Australia Telescope National Facility, PO Box 76, Epping NSW, 1710, Australia}
\offprints{morganti@astron.nl}

\date{Received ...; accepted ...}

\date{\today}

\abstract{We present a detailed study of PKS~1814-637, a rare case of  powerful radio source (P$_{\rm{5 GHz}} = 4.1 \times 10^{25}$~W
Hz$^{-1}$) hosted by a disk galaxy.  Optical images  have been used to model the host galaxy morphology  confirming it to be
  dominated by a strong (and warped) disk component that is observed
  close to edge-on to the line of sight.  This is the first radio galaxy found to reside in a disk dominated galaxy with radio luminosity equivalent to powerful FRII objects. 
At radio wavelengths, PKS~1814-637 is about  480 pc in diameter and it is classified as a compact steep spectrum (CSS)  source; such sources are usually considered to be radio sources observed in the early stages  of their evolution.  However, the optical and mid-IR spectroscopic properties of PKS1814-637
show more in common with Seyfert galaxies than they do with
radio galaxies, with the detection of H$_2$, and PAH emission features, along
with HI and silicate absorption features, providing evidence for a rich ISM which is likely to be related to the disk morphology of the host galaxy. 
We argue that the interaction between the radio plasma and the rich ISM in this  and similar objects may have 
boosted their radio emission, allowing them to more easily enter flux limited samples of radio
sources. In this case, PKS~1814-637  represents a type of
{\sl ''imposter''}: an intrinsically low power object that is
selected in a radio flux limited sample because of the unusually efficient conversion
of jet power into radio emission. This would make PKS~1814-637 an
extreme example of the effects of jet-cloud interactions in galaxies 
containing a rich ISM, and perhaps a missing link between radio
galaxies and radio-loud Seyfert galaxies.  However, it is unlikely that jet-cloud  interactions alone can account for the unusually powerful radio emission compared to Seyfert galaxies,  and 
it is probable that the jet in PKS~1814-637 is also intrinsically  more powerful than in typical Seyfert galaxies, perhaps due to a higher bulge and  black hole mass. The estimated BH mass is indeed higher than the majority of Seyfert galaxies in the local Universe. We speculate that sources similar to PKS1814-637 are likely to be more common at high redshifts, because of the enhanced probability at earlier epochs of triggering radio sources in moderately massive bulges that are also gas-rich.}
\keywords{galaxies: active - galaxies: individual: PKS~1814-637 - ISM: jets and outflow - radio lines: galaxies}
\maketitle  

\section{Introduction}
\label{sec:introduction}

The origin of  the dichotomy  between radio-quiet and radio-loud
objects is still a matter of debate: is it a question of {\it nature} or of {\it nurture}?  Both intrinsic differences in the central engine and extrinsic differences in the surrounding medium have been considered. It is intriguing that  optical AGN (Seyfert-like type) are {\sl mainly} found in late-type, disk galaxies (both in the local as in the far away Universe, see e.g. \citealt{schawinski10}), while this is not the case for radio-loud AGN. These are {\sl mainly}  hosted by
early-type galaxies (\citealt{best05} and refs therein).

The fact that Seyfert galaxies do have, on average, radio core powers  lower than those of radio galaxies 
suggests that they are weaker radio sources from the  outset (see e.g. \citealt{bick98}) and points, therefore, to  a connection with intrinsic
differences e.g.  BH mass and/or spin. However, the characteristics of
the inter-stellar medium (ISM)  in which a radio source is born appear
also to play a role in its further evolution. 
This suggests, for example,  that entrainment  by the radio jet may
have some influence on the level of kiloparsec-scale radio emission
\citep{bick98} and/or that in radio sources hosted by  spiral
galaxies,  even initially relativistic jets can be rapidly
decelerated by collisions in a dense surrounding broad line region or
ISM (see e.g. \citealt{taylor89}).  Connected to this,  the
possibility that the radio emission could actually be temporarily
enhanced by such interactions,  see e.g. \citep{gopal91}, further
complicates  our classification of radio sources.

Although it is extremely rare for nearby radio galaxies to be hosted by genuine disk galaxies, a few   
examples do exist that have been studied in detail (see Table \ref{tab:list} for a summary).  For example, the spiral galaxy 0313-192 in 
the Abell cluster A428 hosts a large double-lobed Fanaroff-Riley I (FR
I) radio source \citep{ledlow98,ledlow01,keel06}, while NGC~612  is an
S0 galaxy with a powerful FR-I/FR-II hybrid radio source and a
large-scale star-forming \HI\ disk \citep{veroncetty01,emonts08}. B2~0722+30 is another  FR I radio
galaxy whose jets are mis-aligned to the galaxy disk, but appear to be
aligned with an \HI\ bridge towards a nearby, interacting pair of
galaxies  \citep{emonts09}. Other possible examples include 3C~293 
\citep{vanbreugel84} and 3C~305 \citep{heckman82}, whose host galaxy morphologies
are highly disturbed due to recent galaxy interactions, but show disk-like
characteristics. An other very recent example is the intermediate redshift disk-like objects that shows spectacular radio lobe structures on scales of several 100 kpc suggestive of relicts \cite{hota11}.

These rare cases of disk-dominated radio galaxies provide an excellent
opportunity for studying the host galaxy properties and environmental
effects that could be important for the triggering and/or evolution of
their radio sources.  In addition, a detailed knowledge of these
systems  provides valuable information for a comparison with
studies at high redshift. Some of these studies are now in progress,
exploring whether disk-like host galaxies may be much more common at
high  than at low redshift \citep{norris08}. However,  much more
will be done in the near future,  due to planned large radio and optical
surveys. 

\begin{figure*}
\centerline{\psfig{file=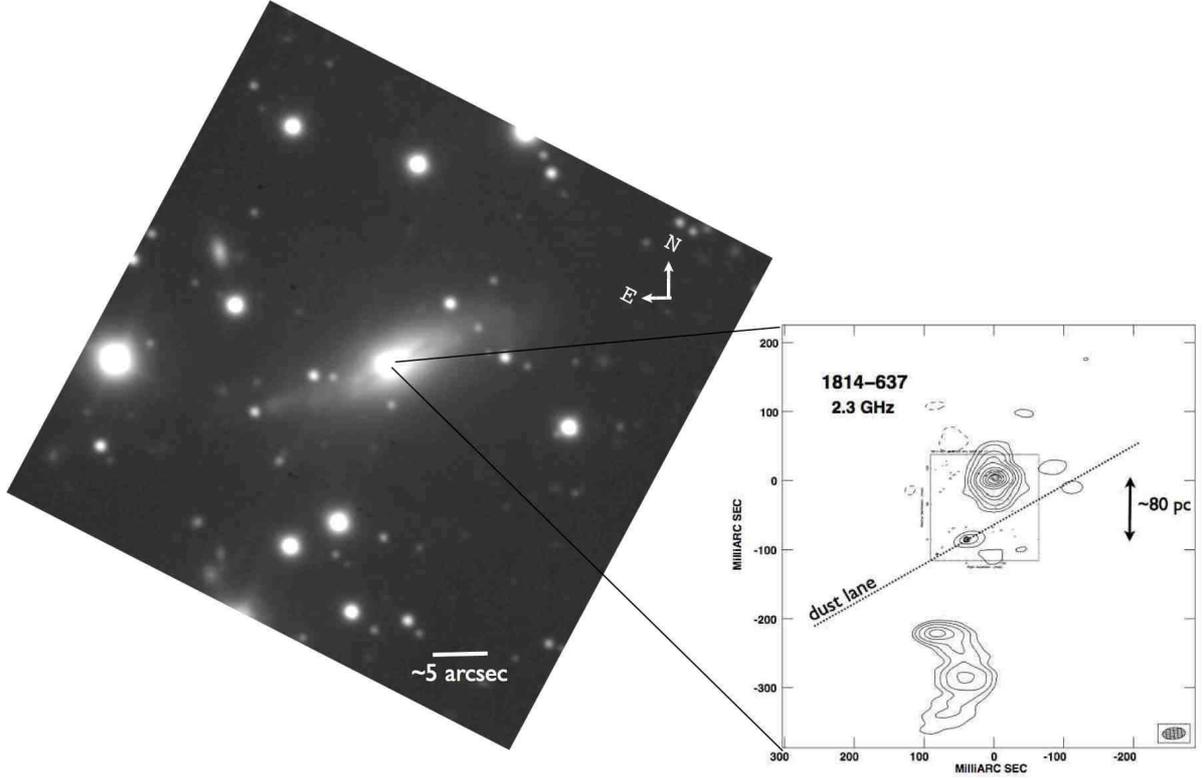,width=16cm,angle=0}}
\caption{{\sl Left}: Optical $r'$-band GMOS-S image of PKS~1814-637 obtained from the
Gemini South. The bright peak coincides with a foreground
star. Despite the presence of the star, the extended disky component and 
dust-lane structure in the galaxy are clearly visible.   {\sl Right}: 
Composite radio map showing 
the 2.3~GHz VLBI image (Tzioumis et al. 2002)  with the 
8.4 GHz  (i.e. high resolution) image from Ojha et
al. 2004) superimposed.  This overlay illustrates the presence of a compact component  (SE of
the brighter lobe) that becomes prominent in the high frequency
observations: we identify this component with the radio core. 
The image at  2.3~GHz 
was obtained from the SHEVE array, the peak a
level is 1.7 Jy and contours are shown at 
-1.5,1.5,3,6,12,18,35,50,65,80 \% of the peak (from Tzioumis et
al. 2002).}
\label{fig:img1814s}
\end{figure*}

In this paper we present a detailed study of a newly found, extreme
example of a radio source hosted by a disk galaxy (see Fig. \ref{fig:img1814s}). Unlike
all  objects mentioned above, the radio power of PKS~1814-637 (P$_{\rm{5 GHz}} = 4.1 \times 10^{25}$  W
Hz$^{-1}$, \citealt{tadhunter93, morganti03, morganti01}) falls well above the radio power boundary between FRI and FRII radio sources
(P$_{\rm{5 GHz}} \sim 10^{25}$ W Hz$^{-1}$; \citealt{fanaroff74}). For
comparison, its radio power is two orders of magnitude higher than the most powerful
radio Seyfert (NGC~1068, P$_{\rm{5 GHz}} \sim 10^{23}$  W Hz$^{-1}$), and more than a factor of four higher than the next most powerful radio source hosted by a disk galaxy (see Table \ref{tab:list}). Interestingly, radio-loud narrow line Seyfert 1 (NLS1) galaxies can reach  radio power comparable to PKS~1814-637 (see e.g. \citealt {foschini11} and refs therein) but because their radio  emission is  likely dominated by a beamed jet emission,  their {\sl intrinsic} jet (and extended  radio lobe powers) might be orders of magnitude lower, making them more similar to Seyferts or low power FRIs.
Because of all this, PKS~1814-637 stands out as a particularly interesting object.

\begin{table}
\begin{tabular}{llll}
\hline
Object &Redshift($z$) &P$_{5~GHz}$ &Reference \\
& &W Hz$^{-1}$ & \\
\hline
NGC~612 &0.0298 &$8\times10^{24}$ &1\\
0313-192 &0.0671 &$9\times10^{23}$ &2,3,4\\
B2~0722+30 &0.0188 &$6\times10^{22}$ &4,5\\
3C293 &0.0450 &$9\times10^{24}$ &6\\
3C305 &0.0416 &$4\times10^{24}$ &7 \\
Speca & 0.1378 & $3\times10^{24*}$  &  8 \\
PKS~1814-637 &0.0641 &$4\times10^{25}$ &9 \\
\hline
\end{tabular}
\caption{The properties of powerful radio radio sources hosted by disk galaxies.
References: 1. \citet{emonts08}; 2. \citet{ledlow01}; 3. \citet{keel06}; 4. NASA
Extragalactic Database; 5. \citet{emonts09}; 6. \citet{vanbreugel84}; 7. 
\citet{heckman82}; 8. \citet{hota11}; 9. this paper. 
$^*$  Total flux derived from NVSS at 1.4~GHz and extrapolated to 5~GHz assuming $\alpha=0.7$.}
\label{tab:list}
\end{table}

From the radio perspective,  PKS~1814-637 is a Compact Steep Spectrum (CSS)
radio source \citep{tzioumis02},  see also Fig.\ 1, right), of about 480 pc linear size\footnote{Throughout this paper
we use a Hubble constant $H_{\rm o}$= 70 km s$^{-1}$ Mpc$^{-1}$ and
$\Omega_\Lambda=0.7$ and $\Omega_{\rm M} = 0.3$. At the distance of PKS~1814-637 this results in 100 mas =
120 pc, \cite{wright06}.}.  The consensus is that the majority of such
sources are young radio sources. However, this class  can also include
cases that are  unable to become  large due to confinement of the radio source by the dense ISM (see e.g. 
\citealt{reynolds97,kunert04,orienti10b} and refs therein).  Finally, \HI\ in absorption with high optical depth has been detected
in PKS~1814-637 \citep{veroncetty00,morganti01} and has motivated the
VLBI follow up that is presented in this paper.  

Here we will use new optical, IR and \HI\ observations of PKS1814-637
to investigate the morphology of the host galaxy, 
study its ISM, and attempt to understand why such a
strong radio source is located in this  unusual host. 

The paper is organised in the following way. In Section 2 we
characterise the optical morphology by analysing  deep Gemini optical
images recently presented in \cite{ramos11} as part of a larger deep
optical imaging survey of the 2~Jy sample.  In Section 3,  we derive
an accurate systemic velocity of the galaxy by  
analysing  previously unpublished near-IR spectroscopic data and by reanalysing our
optical spectroscopy results (Holt et al. 2008). We also include an investigation of the emission line kinematics and discuss the optical spectral classification of the AGN. In  Sec. \ref{sec:spitzer}
we discuss our recent mid-IR Spitzer data (see also \citealt{dicken11})  and in Sec. \ref{sec:radio} we 
discuss new radio VLBI observations obtained with the  Australian Long Baseline Array (LBA). We  bring all the results together 
in Section 6 and we further discuss the implication for high-$z$ observations in Sec. 7.

\section{Host galaxy morphology}

\label{sec:modelling}
\begin{figure*}
\centerline{
\begin{tabular}{ccc}
\hspace*{-1.5cm}\psfig{file=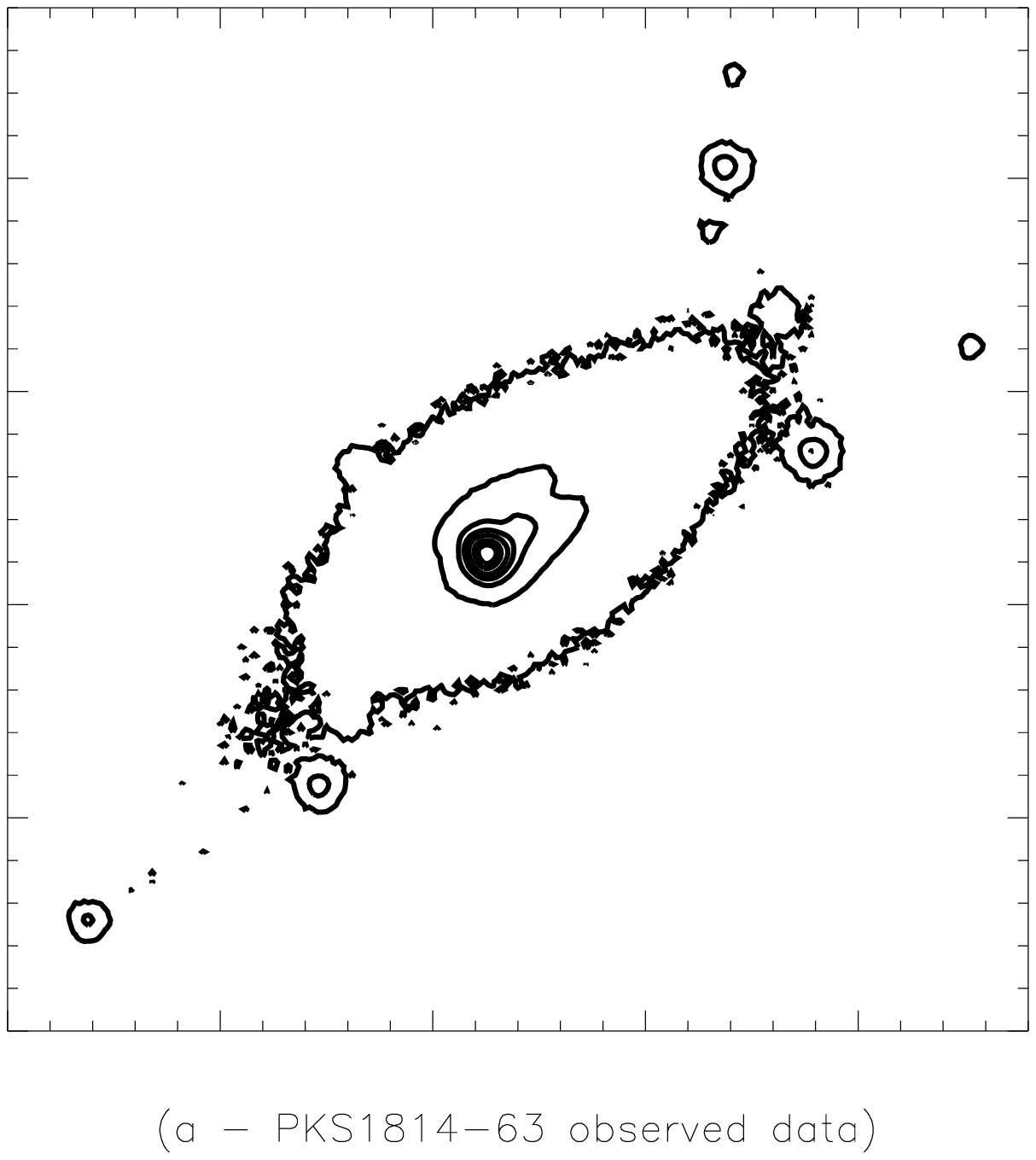,width=6cm,angle=0} &
\hspace*{-1.5cm}\psfig{file=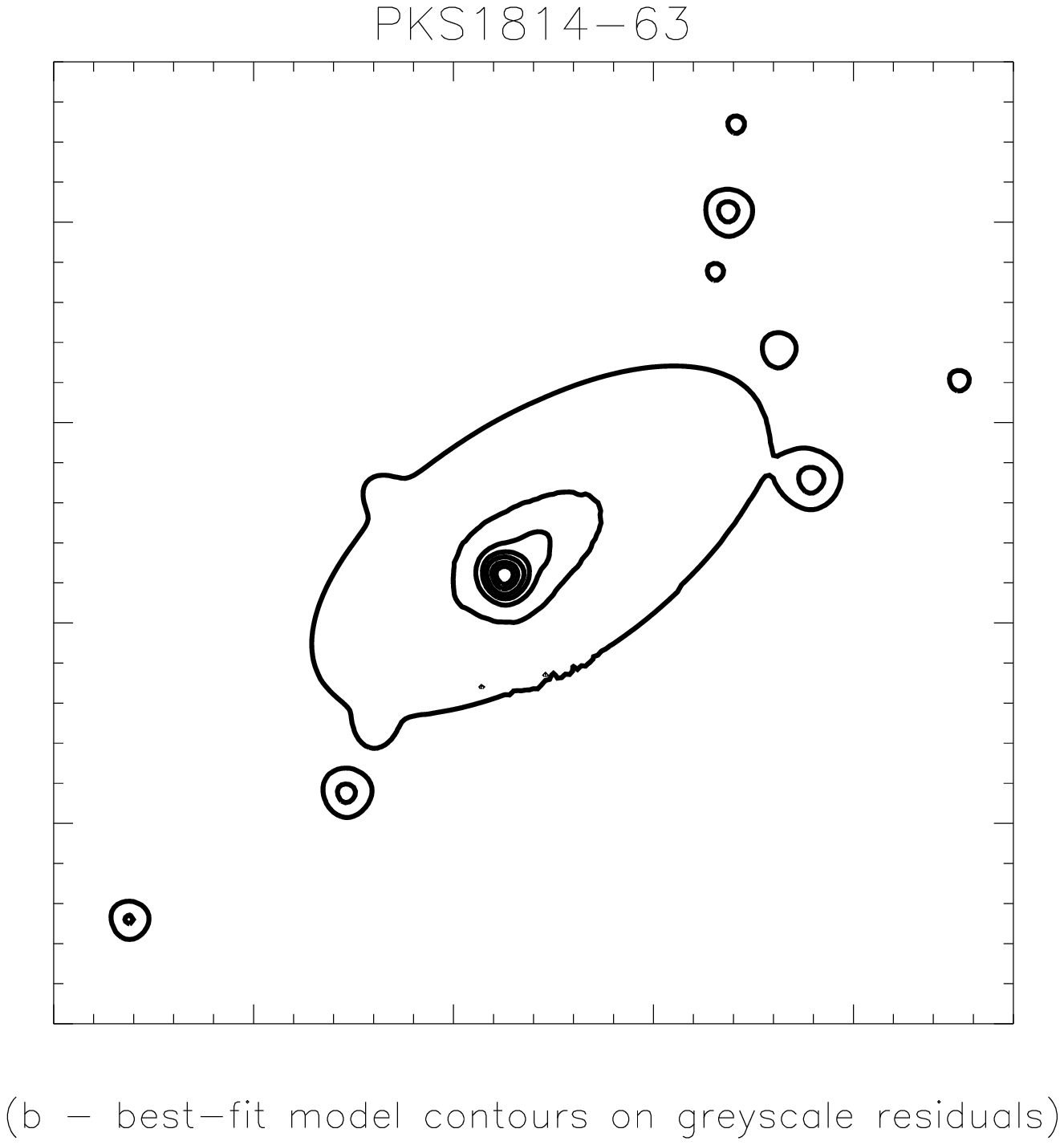,width=6cm,angle=0} &
\hspace*{-1.5cm}\psfig{file=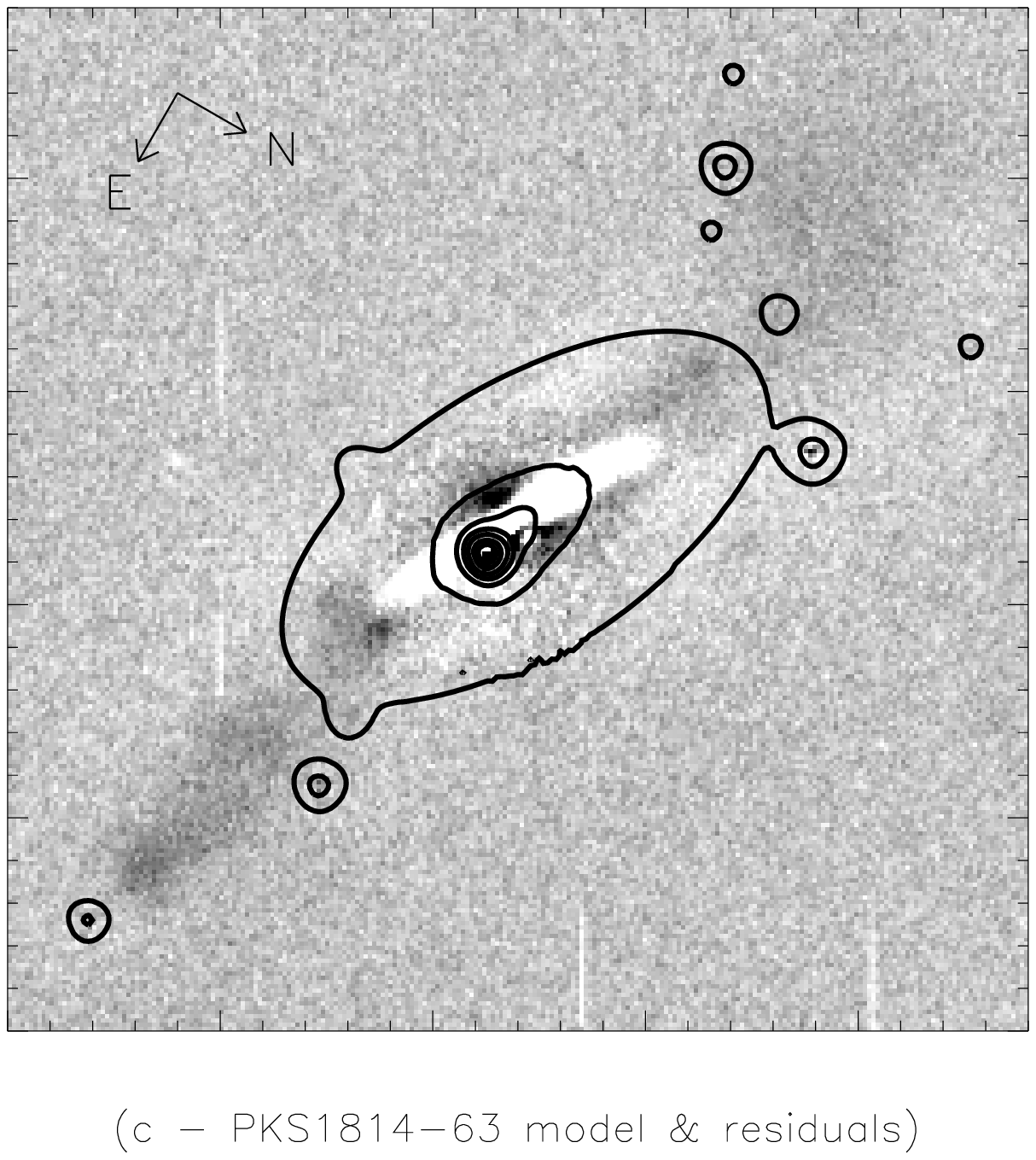,width=6cm,angle=0} \\
(a) & (b) & (c) \\
\end{tabular}
}
\caption{{\sc galfit} modelling results. (a) Gemini  Optical $r^{\prime}$-band  GMOS-S image contours of PKS1814-637, (b) best fitting 
{\sc galfit} model and (c) residuals. See Section 2 for details. Note that the orientation of these images differs from that of the image displayed in Fig. \ref{fig:img1814s}.}
\label{fig:galfit}
\end{figure*}

Although the morphology of the host galaxy of PKS~1814-637  already
appears to be   exceptional  when compared to other radio galaxies
(c.f. deep imaging of the 2Jy sample presented in \citealt{ramos11}),  we now quantify this via modelling  the optical morphology.

Deep optical broad ($r'$) band images of PKS~1814-637 were obtained using the
Gemini Multi-Object Spectrograph South (GMOS-S) on the 8.1-m Gemini South telescope at Cerro Pach{\'{o}}n, Chile (see Figure 1). For 
a full discussion of the observations and data reduction procedure we refer readers to  \citet{ramos11}. 

We have modelled the optical image using {\sc galfit} (\citealt{peng02,peng10}; 
version 3.0)\footnote{{\sc galfit} is a well-documented two-dimensional fitting algorithm which allows 
the user to simultaneously fit a galaxy image with an arbitrary number 
of different model components, in order to extract structural parameters of the galaxy. 
The model galaxy is convolved with a point spread function (PSF) and, using the 
downhill-gradient Levenberg-Marquardt algorithm, is matched to the observational 
data via the minimization of the $\chi^2$ statistic.}. 
This has been a challenging process because of the dust lane, as well as the presence
of the bright foreground star 1.53 arcsec in projection from the 
nucleus of the radio galaxy.  Shorter
exposure time images were used in the modelling in order to avoid saturation effects. We derived a
PSF profile by extracting 2D images of stars in the   
GMOS-S image, normalizing to unit flux and taking an average profile weighted 
by the signal-to-noise ratio of the component extracted stellar profiles. 
The host galaxy was modelled over a 84$\times$84 kpc$^2$ area using 
a S\'ersic profile \citep{sersic63}, and two Gaussian components for fitting the 
foreground star and the unresolved nuclear point source emission from the AGN. 
All model parameters, including the host galaxy, star and AGN centroids 
were allowed to vary freely. We also left the residual background 
level as an additional free parameter. In order to obtain a reasonable
fit, it was necessary to mask the dust lane.
Since the galaxy is in a crowded field, we also iteratively modelled all  neighbouring stars 
and galaxies that interfere with the host galaxy model fit. Once a good model for these adjacent 
objects had been obtained, their parameters were held fixed, effectively removing them from consideration. 
The final reduced-$\chi^2$ value was determined by repeat  modelling 
all other objects in the field of view, resulting in an ideal value of 1.053.

The best fit to the optical image of PKS 1814-637 is a  S\'ersic profile with index $n=2$ which is shown in Figure
\ref{fig:galfit}. Models using more elliptical profiles (e.g. S\'ersic $n=4$) fail to fit the optical image of PKS~1814-637. The estimated
effective radius of the single $n=2$ model is $R_{eff}$=6.4 kpc,
and the position angle of its major axis $PA=-57\degr$, ellipticity $b/a=0.48$,
and  magnitude 
$r^{\prime}=15.75$ mag. The AGN magnitude was found to be $r^{\prime}$=20.98 mag and that of the 
foreground star $r^{\prime} = 15.47$ mag. 
Other models were tried, such as excluding the very faint AGN component and
a two-component, S\'ersic fit. None of these fits gave a better fit to
the data and so the simplest model providing a good fit is preferred. 
Figure \ref{fig:galfit} shows the contour plots of only the central 50$\times$50 kpc$^2$  of 
the modelled region, together with contours of the best-fitting model overlaid on a grey-scale 
of the model-subtracted residual images in the same region.  The residuals of the fit clearly show the central dust lane, aligned with a disk of 
$\sim$45 kpc in length, which is heavily warped at the extremes; the fact that the dust lane intercepts the bulge of the galaxy close to the position of the nucleus implies that the inner disk of the galaxy is observed close to edge-on ($i > 80\degr$).

To summarise, {\sc galfit} modelling  is consistent with PKS 1814-637 having a strong disk morphology, confirming the results based on
visual examination of the images. In addition, the heavily warped outer parts of the disk provide evidence
that the galaxy has undergone a recent galaxy interaction, perhaps related to the triggering of the
activity.

\begin{figure}
\centerline{\psfig{file=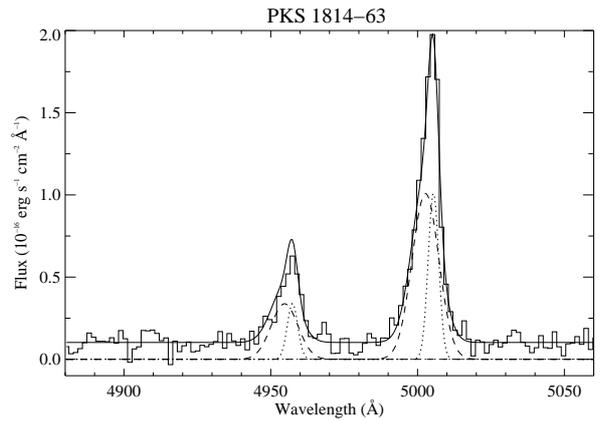,width=8cm,angle=0.}}
\caption[Nuclear \OIII\ line profile.]{Model to the
\OIII$\lambda\lambda$4959,5007 emission-line doublet in the nuclear aperture. The solid lines represent the 
star-subtracted spectrum and the  overall  model fit to the doublet. The components are: narrow (dotted) and
broad (dashed). The two components are overplotted on the radial
velocity profile in Figure \ref{fig:kinematics}. }
\label{fig:o3}
\end{figure}

\section{Re-deriving the systemic velocity } 

\label{sec:optical}

In order to understand the role of the various gaseous components
in PKS 1814-63, it is essential to determine an accurate
systemic velocity. Like the morphological study, spectroscopic investigation of PKS1814-637 is complicated by
the presence of the bright foreground star close to the nucleus of
the galaxy. As shown in \cite{holt08}, the nuclear emission line profiles
can be modelled using two Gaussian components (see also Figure 3). The
narrowest component is spectroscopically unresolved, and the broader
component has a FWHM of 569 $\pm$ 35 km s$^{-1}$. 

\begin{figure}
\centerline{\hspace*{-0.5cm}\psfig{file=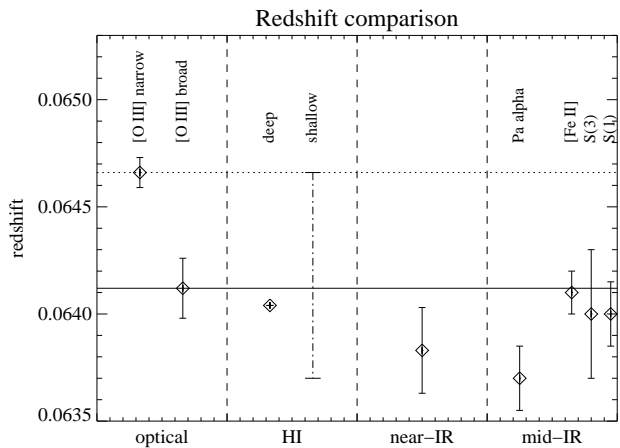,width=9cm,angle=0}}
\caption[Redshift comparison.]{Comparison of the redshifts derived
from the various observations. The horizontal lines represent the
original (dotted; from \protect\citealt{holt08}) and new (solid; this paper) 
interpretations of the systemic velocity. The vertical dot-dashed line
labelled `shallow' marks the range of velocities over which the
shallow \HI\ absorption is observed \protect\citep{morganti01}. The data are also summarised
in Table 1.} 
\label{fig:kinematics}
\end{figure}

As for many other sources in their compact radio sources sample, \cite{holt08} assumed that the optical narrow component 
in PKS1814-637 represents
 the systemic velocity. However, this result has always been considered uncertain: only half of the optical
 radial velocity profile (or `rotation curve') is observed due to the presence of 
 the bright foreground star, which completely dominates the flux on
 one side of the galaxy (see Figure 1 of \citealt{holt08} and Figure B9 of \citealt{holt09}). Here we re-analyse the
 optical spectroscopic data with the help of other available data.  

Figure \ref{fig:kinematics}  shows the redshifts derived from the  two
optical components from  \cite{holt08} and from the deep and shallow
\HI\ components from  \citet{morganti01}, see also
Sec. \ref{sec:radio} for more details. Also plotted are the 
redshifts derived from new Spitzer data (see  Sec. \ref{sec:spitzer}) and near-
IR data from NTT/SOFI (M. Bellamy, priv. comm.). 
Combining the new data and the original optical data, it is clear
from Fig. \ref{fig:kinematics} that a more 
likely interpretation of the kinematics is that the detected optical
broad component is consistent with the systemic velocity, as 
this is also consistent with all other measured redshifts. While
the optical broad component has a relatively large FWHM, this can
be explained in terms of unresolved rotation in the inner disk. 
In this interpretation,
the optical narrow component is associated with the (quiescent) disk of the galaxy
rotating away from the observer on the west side of the galaxy; the rotation
of the large-scale disk is
not clearly detected on the east side of the nucleus due to the
bright foreground star, but it is detected in the nuclear aperture along with the
broad component (see Figure 3). The failure to detect a blueshifted narrow component
in the nuclear aperture (corresponding to the part of the extended disk rotating towards us) 
is likely to be due to a combination of uneven gas emission and dust obscuration. 

 Hence, in our new kinematic interpretation, no nuclear outflow is observed
in this source, making it unusual amongst compact radio
sources (c.f. \cite{holt08}). The new, heliocentric corrected systemic redshift of
PKS 1814-637 is 0.06412$\pm$0.00014. The kinematic offsets of the various
components have been calculated with respect to this new systemic
redshift and are presented in Table  \ref{tab:kinematics}.

\begin{figure}
\centerline{\psfig{figure=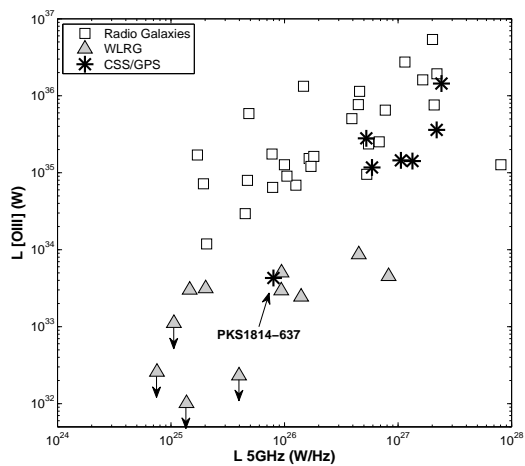,width=7cm,angle=0}}
\caption{Plot of  \OIII\ luminosity versus 5~GHz radio power. The points represent radio sources from the 2Jy sample  (Dicken  et al. 2011), with the CSS/GPS objects marked as stars. } 
\label{fig:o3_rad}
\end{figure}

\section{Optical spectral classification}

Recently there has been much speculation that the optical spectral classification of a radio galaxy is strongly correlated to the rate or mode of accretion of material onto its central supermassive black hole. Radio galaxies with strong optical emission lines (also labelled as  ``high excitation galaxies'', HEGs), including narrow line radio galaxies (NLRG), broad line radio galaxies (BLRG) and quasars, are thought to be energised by the accretion of cool/warm material via a thin accretion disk. On the hand, weak line radio galaxies (sometimes labelled as ``low excitation galaxies'', LEGs) may be powered by the Bondi accretion of the hot phase of the ISM. Details about  this classification can be found e.g. in  \cite{hardcastle06,buttiglione10}. In this context, it is interesting to consider the optical spectral classification of PKS~1814-637. 

In terms of its emission line luminosity, PKS~1814-637 falls (by an order of magnitude) well below 
the correlation between emission line luminosity and radio power (see Fig.  \ref{fig:o3_rad}). In this sense
it is similar to the WLRG, that are defined to
have small [OIII]$\lambda$5007 emission line equivalent widths ($EW_{[OIII]} < 10$ \AA, see 
Tadhunter et al. 1998).  However, in contrast to the other WLRG in the 2Jy sample, our previous spectroscopic
investigation of PKS~1814-637 \citep[see][]{holt08} suggested a  higher equivalent width and classification as a NLRG. 

Unfortunately, our previous estimate of the equivalent 
width derived from the long-slit spectrum was potentially hampered by the (uncertain) subtraction of the continuum associated with the
bright star near the nucleus. Therefore we have re-estimated it using better data. Our Gemini images, which have good seeing, allow more reliable
subtraction of the star and determination of the galaxy continuum flux
in the aperture used for the spectroscopic observations. 
They confirm that the nuclear continuum level is similar to that in the spectrum presented in \cite{holt08}. Combining this information
with the most reliable estimate of the \OIII\ emission line flux (Tadhunter et al. 1993), we find that PKS~1814-637 has an \OIII\ equivalent width in the range 50
- 100\AA, whereas we define WLRG to have EW$_{\rm [OIII]} <$ 10\AA. Therefore this
object is truly ambiguous: it appears like a NLRG in terms of \OIII\ EW, but
more like a WLRG in terms of the [OIII] emission line luminosity;  
PKS~1814-637 is classified as a NLRG, despite its low L$_{\rm [OIII]}$, because it
has an unusually low (stellar) continuum flux. We can
naturally link this with the unusual morphology of the host
galaxy: a disk galaxy with a relatively low central surface brightness, rather
than an elliptical with a high central surface brightness.

It is interesting that the emission line luminosity of PKS~1814-637 
(and indeed other WLRG) falls within the range measured for Seyfert
galaxies in the local Universe. This leads to the intriguing
conclusion that {\sl if it were not for the radio data,
PKS~1814-637 would have been classified as a typical Seyfert galaxy in
terms of its optical and mid-IR spectra (see below), and optical
morphology}. 

\begin{table*}
\caption[Gas kinematics.] {Summary of the kinematic data. Columns are:
  (1) measured emission/absorption line, (2) emission/absorption line
  component, (3 \& 4) velocity width (FWHM) and error in km s$^{-1}$, (5 \& 6)
  velocity shift and error (km s$^{-1}$) with respect to the assumed
  systemic velocity (taken to be the nuclear broad component of {[O
      III]}; see Section \ref{sec:optical}), (7 \& 8) redshift and error with respect to
  the broad component of {[O III]} and (9) references: H08:
  \protect\citet{holt08}; M01: \protect\citet{morganti01}; D11: Dicken
  et al. 2011; V0: \protect\citet{veroncetty00}.  $^{a}$ Full
  width at zero intensity (FWZI) of the broad, shallow 
absorption. $\dagger$ 
\citet{veroncetty00} give e.g. FWZI and we have estimated
the FWHM from their figures, but also give their data here.} 
\label{tab:kinematics}
\begin{center}
\begin{tabular}{ll rrrrrrr}\hline\hline
Lines & Component & Velocity & $\Delta$ & Velocity & $\Delta$
& $z$ & $\Delta$ & Ref\\
  & &  width & & shift & & &  \\
&& (km s$^{-1}$) & (km s$^{-1}$) &(km s$^{-1}$) &(km s$^{-1}$) & \\
(1) & (2) & (3)& (4)& (5)& (6)& (7) & (8) & (9)\\ \hline
\multicolumn{9}{l}{\bf Optical} \\
 {[O III]} & n & unres & & +162 & 21& 0.06466  & 0.00007 & H08\\
   &b & 569 & 35 & 0 & 20 &0.06412 &0.00014 & H08\\
   & 1 Gaussian & 411 & 17 & & & & &H08\\
\\
\multicolumn{9}{l}{\bf Radio}\\ 
 \HI\ & n & $\sim$50 & & -24 &  &0.06404&&M1 \\
   & b & $\sim$280$^{a}$ & &162 to -119 & & &&M1 \\
   & & 62 $\dagger$ & & -192$\dagger$ & & && V0$\dagger$\\
\\
\multicolumn{9}{l}{\bf Mid-IR} \\ 
 {[Ne II]}, {[Ne III]} & 1 Gaussian & &&-87&60& 0.06383 &0.00020 & D11 \\
   {[O IV]}, H$_{2}$ \\
\\
\multicolumn{9}{l}{\bf near-IR} \\ 
Paschen $\alpha$ & & & & -126 & 60 & 0.06370 & 0.00015 & \\
{[Fe II]} & &&&-6 & 30 & 0.0641 & 0.0001 & \\
S(3) &&&&-36 & 90 & 0.0640 & 0.0003 & \\
s(1) &&&&-36 & 60 & 0.0640 & 0.00015 &\\ \hline\hline
\end{tabular}
\end{center}
\end{table*}
\begin{figure*}
\centerline{\psfig{figure=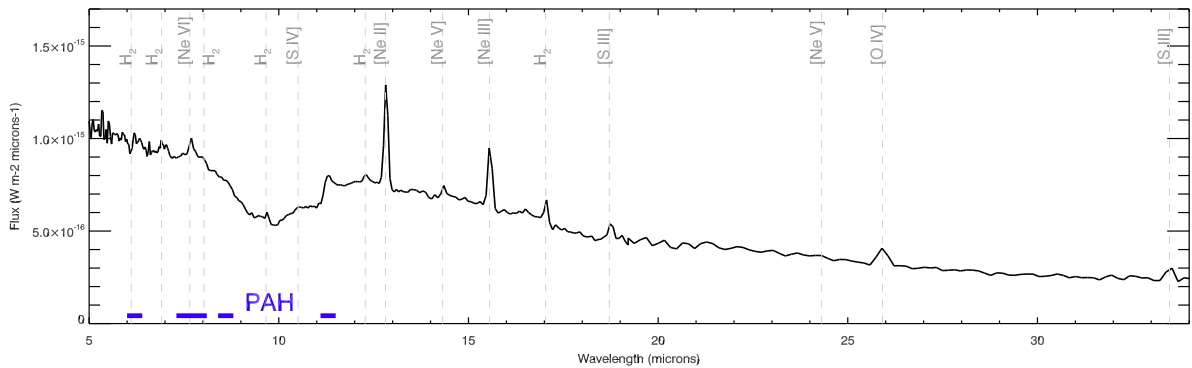,width=16cm,angle=0}}
\caption{{\it Spitzer} Mid-infrared spectrum of PKS~1814-637. Fine
  structure lines are indicated  as well as the position of strong PAH
  emission bands at 6.2, 7.8, 8.6 and 11.3 microns. Note the strong 10 micron silicate absorption feature.} 
\label{fig:PKS1814_Spitzer}
\end{figure*}

\section{The ISM of PKS~1814: Spitzer mid-IR data}
\label{sec:spitzer}

Figure \ref{fig:PKS1814_Spitzer} shows the Spitzer IRS spectrum
obtained as part of our multi-wavelength investigation of the 2~Jy
sample (Dicken et al. 2011). The spectrum shows prominent fine
structure lines, as well as PAH and H$_2$ emission features, and a silicate absorption band.  
The detection of emission lines in  the mid-IR spectra of radio
galaxies is common, and the presence of high ionisation potential lines, such as
the [OIV] and [NeV] lines, clearly detected in PKS1814-63, 
indicates a significant AGN photoionised component (see e.g. \citealt{ogle10}
and refs. therein). However,  the 
detection of H$_2$ and PAH features is rare
in radio galaxies in general --- only 5 objects (12\%) in the 2Jy
sample show significant H$_2$
emission, and only 10 objects (23\%) have detected PAH bands 
\citep{dicken11}. 

The most prominent feature in the mid-IR spectrum of
PKS~1814-637 is the 10\um\ silicate absorption feature. The absorption
depth of the silicate feature in PKS~1814-637 is relatively high  ($\tau_{10\mu m} =  0.48$) 
compared to the 40\% of NLRG in the 2~Jy sample that show a silicate absorption feature in their mid-infrared spectra. 
In the context of orientation based unification
schemes for AGN, the detected silicate absorption is consistent with
an AGN viewed edge-on that is obscured by circum-nuclear
dust i.e. a NLRG. However, the 10\um\ silicate
absorption feature in PKS~1814-637 is unique for CSS  in the 2~Jy sample. This fact may be related to the large-scale dusty disk
component that is more prominent in the case of
PKS~1814-637 than in any other radio galaxy in the 2~Jy sample. 

Overall, the  mid-infrared spectral data for PKS~1814-637 provide
evidence for the presence of a rich ISM, with a range of ionized gas, molecular
and dust features detected. Although strong PAH and H$_2$ features (but not silcate absorption: see above) appear to be a relatively common feature of
CSS sources  \citep{dicken11}, their detection is rare for powerful radio galaxies in general. Indeed, the overall character 
of the mid-IR spectrum of PKS~1814-637 --- with its mix of high and low ionization
fine structure, H$_2$, silicate and PAH features --- shows greater similarity
to Seyfert galaxies \citep{gallimore10,baum10}  than it does to typical radio galaxies \citep{ogle06,dicken11},
as already pointed out  in the previous section. 

\section{The ISM of PKS~1814: radio continuum and 21-cm atomic neutral hydrogen}
\label{sec:radio}

\subsection{Previous radio continuum VLBI observations}

PKS~1814-637 was observed in VLBI at 2.3 GHz by \cite{tzioumis02}. Fig. \ref{fig:img1814s} (right) 
shows the complex structure of the source which has an overall extent of  $\sim 400$ mas (i.e. $\sim 480$ pc).  
Two lobe-like structures with very different morphologies are observed: the southern region is dominated by two components with 
similar brightness embedded in a relatively large diffuse component, while  the northern region shows a prominent bright component  with possible
North-South symmetrical extensions.  Just over 50\% of the total flux density of the source is detected in the VLBI observations, indicating the
presence of a more diffuse extended component (but limited to the
arcsec scale as no extended component was reported from the ATCA
observations).  

A weak radio continuum component is also observed (at the 5\% level)
between the two major lobes in the 2.3~GHz image, which corresponds to
a 15$\sigma$ detection.   Interestingly,  8.4~GHz VLBI observations
\citep{ojha04} confirm the presence and prominence of this component
at high frequency, suggesting this component is likely to be the radio
core in PKS~1814-637  (see Fig.  \ref{fig:img1814s}  right). 

As a final remark, it is worth mentioning that the radio structure is
not perpendicular to the galactic disk and dust lane (see
Fig. \ref{fig:img1814s}) as in many dust lane galaxies (e.g. Cen A),
but forms an angle of about 50$^\circ$ (projected) to the galactic disk and dust
lane.

\begin{figure}
\vspace{-3mm}
\centerline{\psfig{figure=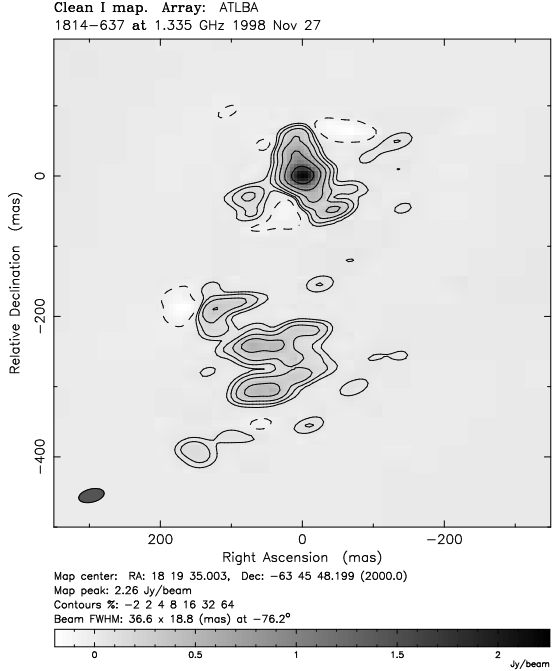,width=8cm,angle=0}}
\vspace{0.3mm}
\caption{PKS~1814-637 at 1335 MHz as obtained from the line-free channels of
the observations presented in this paper. }
\label{fig:img1814_1335}
\end{figure}

\begin{figure}
\centerline{\psfig{figure=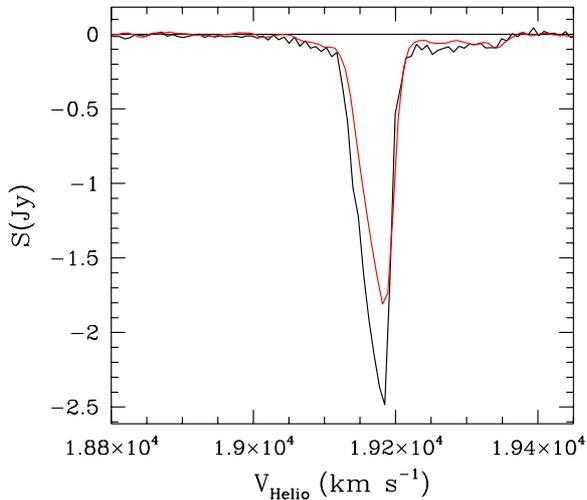,width=8cm,angle=0}}
\vspace{0.3mm}
\caption{Comparison of the ATCA (black) and LBA (red) integrated \HI\ absorption  profiles.}
\label{fig:HIcomparison}
\end{figure}

\subsection{LBA observations: data reduction and radio continuum}

 \HI\  observations centred at 1336~MHz   were obtained with the Australian Long Baseline Array (LBA) on 27 Nov 1998.
The array comprised four stations; Parkes (64 m), Mopra (22 m), the Australia Telescope Compact Array (5$\times$22-m dishes as
tied-array), and the Mount  Pleasant 26-m antenna of the University of Tasmania.  We used a recording band of 8~MHz  width in each
circular polarisation and  256 spectral channels.  
The editing and part of the calibration of the data was done in {\sc aips} and then the data were transferred to {\sc miriad} \citep{sault95} for the bandpass calibration.  The calibration
of the bandpass was done using additional observations of the strong
calibrator PKS~1921--293.  
The resulting velocity resolution is  $\sim 7$ \kms\ before Hanning
smoothing.

The  line  cube was made using uniform weighting after subtracting
the continuum emission from the $uv$-data using the line-free channels. The noise per channel is $\sim 4$ \mJybeam\ after
Hanning smoothing  and the restoring beam size is $33 \times 16$ mas (p.a.\ = $-72.4^{\circ}$).   

A continuum image  was obtained  using the 
line-free channels. The image is shown in Fig. \ref{fig:img1814_1335} and it was obtained using {\sc aips}
and {\sc difmap}. The beam size is  $36 \times 18$ mas (p.a.\ = $- 76^{\circ}$). These data were originally intended to provide information
about the spectral index (in combination with the 2.3~GHz data). However, the data quality  prevented us from  fully achieving this goal. 
The rms noise in the image is $\sim  12$ mJy/beam.  Although the quality is clearly inferior to the 2.3~GHz image of Tzioumis et
al. (2002) -  the rms noise is more than twice as high -  the 1.3~GHz image confirms the overall structure of the
source.  

The total flux detected is  $S_{\rm 1.3~GHz} = 11.5$ Jy. Compared with the 13.5 Jy detected with  ATCA observations it
confirms that a  fraction of the flux (at least 2 Jy,
i.e. 15\%) is likely undetected because it originates in diffuse, low surface
brightness emission.   Because of the calibration problems mentioned above, we could only attempt an  estimate of the
integrated spectral index (using 1.3~GHz and 2.3~GHz images convolved to the
same restoring beam).  We obtain values between 1.3~GHz and 2.3~ GHz of
$\alpha^{1.3}_{2.3} \sim -1.4$  for the southern lobe  and
$\alpha^{1.3}_{2.3} \sim -1.15$ for the northern 
lobe, consistent with our assumption that both structures are in fact radio lobes.
Higher quality data will be necessary for an accurate estimate of the spectral index.

\subsection{Results from the  \HI\ absorption}

\begin{figure*}
\centerline{\psfig{figure=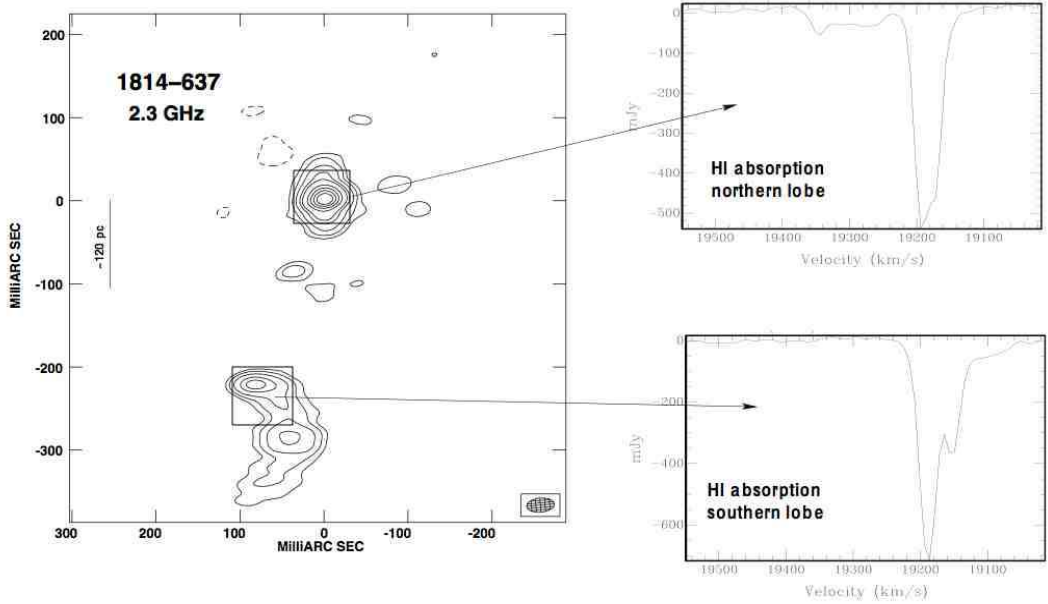,width=14cm,angle=0}}
\caption{VLBI 2.3~GHz image of PKS~1814-637 (left) from
  \cite{tzioumis02}. The locations of the two {\sl integrated}
  \HI\ spectra  shown on the  right are marked. The two spectra clearly illustrate that the deep
  absorption is present at both locations while the  broad, shallow
  absorption changes drastically going from the northern to the
  southern lobe.} 
\label{fig:Panel}
\end{figure*}

The LBA observations show that the \HI\ absorption  is extended and complex on the VLBI scale.
The VLBI \HI\ observations  recover a large fraction  of the absorbed flux observed at low resolution with the ATCA (see Fig. \ref{fig:HIcomparison}).
The ATCA profile already suggested  the presence of at least two
components  of \HI\ absorption: a deep and relatively narrow component
(with high optical depth, $\tau \sim 20$\%) and broad and shallow
wings. Interestingly, these components  have, on the  VLBI scale,
different spatial distributions. As can be seen in
Fig. \ref{fig:Panel},  the deep absorption is extended and covers the
entire source, while the shallow wings are more localised. The
redshifted wing is observed {\sl only} against the northern lobe while
the blueshifted wing is detected against the southern lobe.   The
redshifted wing appears to be more prominent than the blueshifted one.  
The column density is $\sim 3 \times 10^{20}$ cm$^{-2}$ for the deep
component assuming a T$_{spin}$ (temperature that gives the relative population of the two levels) $\sim$ 100 K.

As discussed in Sec. \ref{sec:optical},  the central velocity of the
deep absorption is consistent with the results from the optical and IR emission lines
and is  defined as systemic velocity. Interestingly, the range in
velocity covered by the \HI\  in the broad wings is comparable to the
velocity range seen in the ionised gas (see Fig. \ref{fig:kinematics}
and Table \ref{tab:kinematics}).  

The \HI\ absorption can be interpreted as coming from two separated
gaseous structures, with the  deep absorption due to cold gas
located at large distance from the nucleus and  radio source  likely associated with the large-scale disk of the host
galaxy. This would explain why this component is detected over the
entire source and  peaks around the systemic velocity. Furthermore,
the high optical depth  would be due to gas with a column density
similar to that found in large scale disks, e.g. in Seyfert galaxies
but also for radio-quiet early-type galaxies \citep{gallimore99}, and
characterised by a  T$_{spin}$ $\sim$ 100 K.
Considering that the large-scale, galactic disk is seen edge-on, the
fact that the deep component is projected over the entire VLBI source
would imply a  thickness of the disk of the order of 400 - 500 pc,
 comparable to what is expected for these types of structures in
spiral galaxies  (in particular in the outer regions where they tend
to flare). The shallow, broad  components could, instead,  trace 
a circumnuclear disk located closer to the radio source:   the broad widths of the shallow, shifted absorption features would be due 
to unresolved rotation or projected along the line of sight. 

If this circumnuclear disk has approximately the same orientation as the
large-scale one, the detection of a velocity gradient would then be due to
the misalignment of the radio structure compared to the gaseous
disk. This would explain the observed velocity gradient (redshifted
against the northern lobe  and blueshifted against the southern
lobe). The low optical depth observed even on the VLBI scale  for the
broad component would not be due to effect of filling factor but could
also be due to an higher spin temperature (i.e. T$_{spin} \gta 1000$
K) of the gas in this structure due to the vicinity of the AGN. This would be consistent with what found for other compact sources (see e.g. \cite{holt06}).

\section{Discussion}

\label{sec:discussion}

The analysis of multi-wavelength data (imaging and spectroscopy)  has
revealed that PKS~1814-637 is an intriguing radio galaxy. The
modelling of the optical image has shown  the presence of a prominent
disk component, confirmed by the best fitting  S\'ersic $n=2$ profile.
This is interesting given the high radio power of
PKS~1814-637: this is the first radio galaxy of such power (FRII-like)
found to reside in a disk dominated galaxy, and the only one of this
type in a complete southern sample of powerful radio sources (2~Jy
sample, see e.g. \citealt{ramos11}). This raises the question of how
this disk galaxy succeeded in producing such a powerful radio source, and why
this is not a more common phenomenon.

\subsection{PKS~1814-637:  an "impostor" radio galaxy?}

\label{sec:imposter}

Given the evidence for a rich ISM we have found for PKS~1814-637 as well as its highly distorted
radio morphology, we argue that the interaction of the radio plasma
with the ISM must have had a major impact on the characteristics of
this source. 

In a study of a sample of nearby radio galaxies,  
\cite{tadhunter11}  noted  the high incidence of
CSS/GPS sources  in particular among the objects showing the presence
of a young stellar population component  in their host galaxies. This
has been suggested to be the result of an observational selection
effect where the strong interaction of the radio jets with the rich ISM,
characteristic of  objects resulting from major mergers, may influence
the conversion of the jet power into radio luminosity. Due to the
compression of the magnetic field and the increased density of particles,
the radio luminosity will be boosted. This would make these objects
more likely to enter a radio flux selected sample of objects.  
Indeed, the possibility of a variation of efficiency with which beam power is
converted into radio emission has been also suggested by
\cite{gopal91} in order to explain the dependence of  linear size of
powerful radio sources on redshift. 

Thus, PKS~1814-637 could represent one of the best examples supporting
this scenario and be an {\sl "imposter"} in the 2~Jy sample: an
intrinsically lower power object that is selected in the sample because
of the rich ISM that may contribute to the efficient conversion of jet power
into radio emission. 

This idea is also supported by the fact that, although PKS~1814-637 is
classified as a NLRG on the basis of the optical spectrum (see
Sec. \ref{sec:optical}), its \OIII\  luminosity is lower than one would
expect on the basis of its radio power \citep{morganti97,holt09}. In Figure \ref{fig:o3_rad} we plot
\OIII\ emission line luminosity versus 5~GHz radio luminosity for the 2~Jy sample, where
$L_{\rm [OIII]}$ is a known to be a good indicator of AGN power. This shows 
that, on average, CSS/GPS sources
lie below the the distribution of radio galaxies for any given
 $L_{\rm Radio}$\footnote{Similar results are found correlating $L_{\rm Radio}$ with 
 other AGN indicators in the mid-infrared i.e. \OIV\ 25.89\um\ and 24\um\ luminosity.}, providing evidence that these objects have enhanced radio
emission for their AGN power. It is notable that PKS~1814-637 has the lowest
\OIII\ luminosity out of all the objects classified as NLRG -- including the other CSS/GPS sources -- in the 2~Jy
sample.

As discussed in Section \ref{sec:optical}, PKS~1814-637 has an emission line luminosity, but not equivalent 
width, similar to that of WLRG, as seen in Figure \ref{fig:o3_rad}. WLRG have also been shown 
to have intrinsically weak AGN for their radio power, and the properties of their local ISM 
are also hypothesised to boost the radio luminosity \citep{dicken09}.

In this context, it is interesting to
note that  large \HI\ disks in a sample of radio galaxies have been
found {\sl only} around compact radio sources
\citep{emonts07}. PKS~1814-637 could be another example in this trend,
considering that the characteristics of the deep \HI\ absorption
suggests that it originates  from a large scale disk (not seen in
emission because the object is too far away and the sensitivity of the
ATCA is too low). This suggests that  
there exists a group of compact radio sources with  large
reservoirs of cool gas, either originating  from gas-rich major mergers or
present in pre-existing disks,
where  the radio emission has been
boosted, at least temporarily,  by the interaction of the radio plasma
with the rich ISM.  Therefore, as the jets expand beyond the central regions of the galaxy that contains the rich ISM,
their radio luminosities will decline until they drop below the flux limit of the particular radio sample being considered. 
Such a scenario leads to a short life time and agrees with the apparent rarity of objects like PKS~1814-637.
In addition to this, because of the many characteristics that we have 
identified in common between PKS~1814-637 and  Seyfert galaxies, this
group of objects may even represent a "missing link" between radio
galaxies and Seyferts.

\subsection{The lack of outflow}

If the conclusion about the systemic velocity is confirmed, it means that, unlike the majority of high radio power  CSS/GPS \citep{holt08},  PKS~1814-637 does not show evidence
for a fast outflow of ionised or atomic neutral gas. This may seem surprising if, as suggested in the previous section, the radio emission is really boosted by strong jet-cloud interactions.

The orientation of the source may help to explain the lack of evidence for an outflow: a relation between the orientation of the source and the amplitude of the outflow has been presented in \citealt{holt09} (i.e. higher inclination lower outflow velocity).  Being close to the plane of the sky, it could be  more difficult to see any outflow produced along the radio axis in PKS~1814-63. However, if this is the case, we might expect to see strong line broadening effects at the site of the jet-cloud interaction (e.g. due to expansion of the radio lobes perpendicular to the radio jets).   It is also  possible that the emission line outflow (but not a neutral outflow) could be hidden by the near edge-on dusty disk at optical wavelengths. In this case the emission lines would represent more extended material that is illuminated by the AGN but not directly involved in the outflow. 

One possible solution may be that the shock induced by the radio jets is so strong, and the gas heated to such high temperatures, that it does not have the chance to cool and to radiate in optical emission lines or absorb in \HI. Therefore,  we only see strong emission from the shock photoionized precursor gas. This is what we think may be happening in Coma A to the SE of the nucleus \citep{solorzano03}, where the jet appears to make a "direct hit" on a dense cloud (and the radio source is clearly deflected at this location), but despite the strong high ionization line emission associated with the cloud, we do not see {\sl any} sign of kinematic disturbance in the emission line kinematics. 
If this is the case, one may expect to detect X-ray (bremsstrahlung) emission. New Chandra observations have been recently obtained and show a soft excess (Mingo B. priv. communications). The lack of adequate  spatial resolution will likely make difficult to confirm whether this excess can be due to bremsstrahlung emission.

\subsection{Link to Seyfert galaxies}

We have pointed out throughout  this paper that similarities exist  between PKS~1814-637 and Seyfert galaxies. 
We also remarked on the basis of our optical data in Sec. \ref{sec:optical}, as well as from our mid-IR data (Sec. \ref{sec:spitzer}),  that without the information from the radio data,  PKS~1814-637 would have been likely classified  as a  typical Seyfert galaxy in terms of its optical and mid-IR spectra, and optical morphology.

However, PKS~1814-637 has much higher radio power compared to even the most powerful Seyfert galaxies.  In the case of radio-loud NLS1, that have been mentioned in the introduction, their radio power comparable to PKS~1814-637 (see e.g. \citealt {yuan08,foschini11} and refs therein) is likely due to their radio  emission is  dominated by a beamed jet emission.  Furthermore,  in general the host galaxies of radio-loud NLS1 have not been yet well characterised and, therefore, it is not clear that these object are host by disk  galaxies. Two examples (see \citealt{zhou07,foschini11}) suggest that this could be the case at least for some of them. If this is confirmed for a larger sample, it may indicate a possible, interesting link between this group of object and objects like PKS~1814-637.

However, if we consider unbeamed, nearby Seyfert galaxies,  even in those where  clear evidence of jet-cloud interactions have been found, the radio emission does not reach the level observed in PKS~1814-637. An example is IC~5063, a radio-loud Seyfert galaxy where a strong ongoing interaction is observed between the radio plasma and the ISM (\citealt{oosterloo00, morganti07} and ref. therein). The radio emission is aligned with the dust-lane, and  the lobe interacting with a cloud of molecular gas  is also much  brighter in radio \citep{oosterloo00}. This object may represent another example of  radio emission boosted by jet-cloud interaction, however the total radio power is almost two orders of magnitude lower than in PKS~1814-637. 

Thus, it is unlikely that the difference between the radio power of Seyfert galaxies and that of PKS~1814-637  is  solely due to the interaction; it is probable that the jet in PKS~1814-637 is intrinsically  more powerful than in typical Seyfert galaxies, perhaps related to a higher bulge and  black hole mass. To investigate this, we have attempted to estimate the mass of the black hole in PKS~1814-637 using different methods. 

First, we have used the velocity gradient observed in the shallow \HI\ component -- that we have identified with a possible circumnuclear disk -- as a probe of the BH mass.   Ignoring projection effects that cannot be quantified, the \HI\ velocities range from  $+180$ to $-100$ km/s relative to the systemic velocity. These velocities are measured at the location of the VLBI lobes, i.e. at a projected distance of about 100 pc from the core. Under these conditions, we estimate a relatively high value for the BH mass in the range between $3 \times 10^8$ and $10^9$\msun.

Alternatively, we can also make use of known BH - bulge mass relations (\cite{magorrian98}) and derive the BH mass from the bulge mass obtained from the K-band images \citep{inskip10} and from the modelling of the galaxy. 
We have used the  K-band magnitude from \cite{inskip10} to obtain a proxy and an  upper limit to the bulge mass. The  K-band magnitude (K$=12.52$) is equivalent to an observed flux of $3.77\times 10^{-16}$ erg cm$^{-2}$ s$^{-1}$ A$^{-1}$. Assuming a galaxy with a 12~Gyr old stellar population, the SED proposed by \cite{maraston09}  and the luminosity distance of $\sim$ 265  Mpc, the observed flux matches a galaxy mass of $\sim 3\times 10^{11}$ \msun. Using this mass and the correlation shown in Fig. 2 of \cite{haring04}, we derive a BH mass again in the range $\sim 3 \times 10^8$ to $\sim 10^9$\msun, nicely consistent with that we derived from the \HI. A similar value of  $6 \times 10^8$\msun\ is obtained by following instead \cite{marconi03}. These black hole mass estimates for PKS~1814-637 are higher than those estimated for any of the nearby Seyfert galaxies in the sample of \citet{mclure01, mclure02}, once corrected to our cosmology.


Finally,  from the modelling of the optical images (see Sec. \ref{sec:modelling}) we can derive an estimate of the bulge mass inside the derived effective radius  ($6.45 \pm 0.01$ kpc) using the integrated bulge magnitude ($r'$-band = 15.75 mag), and thus estimate a lower limit to the BH mass. We can use the absolute magnitude derived in this way (M$_r = -21.52$) to extract, from the correlation shown in  \cite{mclure01}, the BH mass. 
Again, the bulge luminosity of PKS~1814-637 appears to be located at the higher end of the distribution of Seyfert galaxies confirming that, unlike a typical Seyfert galaxy, PKS~1814-637 is hosted by a more early-type galaxy (S0-like), i.e. a galaxy with a large bulge.
Following the prediction of \cite{magorrian98} (see also  \citealt{mclure01} for details) the BH mass would be just below $\sim 10^8$\msun. However the plot in \cite{mclure01} shows a large scatter in the value derived for Seyfert galaxies with comparable bulge mass, up to a few times 10$^8$\msun.  

Thus, PKS~1814-637 appears to host a quite massive BH that would compare only with the most massive BH found in Seyfert galaxies. 
The combination of the strong interaction with the surround ISM discussed in the previous session and  massive black hole (related to the large bulge of the host galaxy) could provide the right conditions to host a powerful radio AGN in this galaxy.

\section{Radio sources like PKS~1814-637: how rare?}

\label{sec:rare}

The case of PKS~1814-637 is particularly interesting because it is the first case of  a {\sl powerful} radio source  in a disk galaxy.
As described above, this allows us to understand the conditions in which a powerful radio source can  be triggered even when hosted by a disk-like galaxy. In addition to this, it also  helps our understanding of whether this type of AGN and radio source were more common in the earlier Universe.
The idea of the interaction between radio plasma and dense ISM affecting the radio emission has indeed been proposed for high-$z$ radio galaxies to explain e.g. the correlation between the steepness of the spectral index and the redshift of the sources \citep{athreya98, klamer06}. 
The possibility of a higher incidence  of powerful radio sources associated with disk galaxies at high-$z$ has been brought up by \citet{norris08}, although they have so far identified only one possible candidate in their deep field. More recently, \cite{schawinski10}  have studied the optical morphologies of host galaxies of AGN at  $1.5<z <3$ X-ray selected (10$^{42} < L_X < 10^{44}$ erg s$^{-1}$) in the CDF-S. The majority of these AGN are hosted by disk galaxies consistent  with what is found for Seyfert galaxies in the local Universe (see also \citealt{cisternas11}). 

Using the deep VLA radio observation presented by \cite{manieri08}, we find that two of the AGN selected by \cite{schawinski10} appear  to have radio counterparts of 3.7 and 0.07 mJy  at 1.4~GHz. These sources are both at $z \sim 1.6$ and converting this to radio luminosity, these sources  
have a radio power of log P$_{\rm 1.4GHz} = 25.6$ W/Hz and 24.1 W/Hz respectively. Considering the limited coverage of this deep field (at most 50 sq armin) this represents an interesting result that may indicate that indeed high power radio sources hosted in disk galaxies could be more common at higher redshift. 

The availability of large  surveys coming on-line in the near future from new radio facilities (e.g. LOFAR, ASKAP, Apertif) should allow the existence of this population of sources to be confirmed.

\begin{acknowledgements}

CRA acknowledges Christian Leipski for very useful comments on the GALFIT fitting.
The Australia Telescope Compact Array (/Parkes telescope/Mopra telescope/Long Baseline array) is part of the Australia  Telescope which is funded by the Commonwealth of Australia for operation as a National Facility managed by CSIRO. 
The authors would like to thanks the referee, Luigi Foschini, for his very  constructive comments.
This research has made use of the NASA Extragalactic Database (NED), whose contributions to this paper are gratefully acknowledged.  This publication makes use of data products from the Two Micron All Sky Survey, which is a joint project of the University of Massachusetts and the Infrared Processing and Analysis Center/California Institute of Technology, funded by the National Aeronautics and Space Administration and the National Science Foundation.
\end{acknowledgements}

\bibliographystyle{aa}
\bibliography{abbrev,refs}
\end{document}